# Uplink Contention Based SCMA for 5G Radio Access

Kelvin Au, Liqing Zhang, Hosein Nikopour, Eric Yi, Alireza Bayesteh,
Usa Vilaipornsawai, Jianglei Ma, and Peiying Zhu
Huawei Technologies Canada Co., LTD.
Ottawa, Ontario, Canada
{kelvin.au, liqing.zhang, hosein.nikopour, zhihang.yi, alireza.bayestech,
usa.vilaipornsawai, jianglei.ma, peiying.zhu}@huawei.com

*Abstract*—Fifth generation (5G) wireless networks are expected to support very diverse applications and terminals. Massive connectivity with a large number of devices is an important requirement for 5G networks. Current LTE system is not able to efficiently support massive connectivity, especially on the uplink (UL). Among the issues arise due to massive connectivity is the cost of signaling overhead and latency. In this paper, an uplink contention-based sparse code multiple access (SCMA) design is proposed as a solution. First, the system design aspects of the proposed multiple-access scheme are described. The SCMA parameters can be adjusted to provide different levels of overloading, thus suitable to meet the diverse traffic connectivity requirements. In addition, the system-level evaluations of a small packet application scenario are provided for contention-based UL SCMA. SCMA is compared to OFDMA in terms of connectivity and drop rate under a tight latency requirement. The simulation results demonstrate that contention-based SCMA can provide around 2.8 times gain over contention-based OFDMA in terms of supported active users. The uplink contention-based SCMA scheme can be a promising technology for 5G wireless networks for data transmission with low signaling overhead, low delay, and support of massive connectivity.

## I. INTRODUCTION

The advent of smartphones and tablets over the past several years has resulted in an explosive growth of data traffic over the cellular network not seen in previous generations. With the proliferation of more smart terminals communicating with servers and each other via broadband wireless networks, new applications emerge to take advantage of wireless connectivity. Fifth generation (5G) wireless networks are expected to support very diverse services – from very low latency to very high delay tolerant, and from very small to very large packets in different applications [1]. An important 5G requirement is that it should support massive connectivity [1] with a large number of devices such as smart-phones, tablets and machines. As a result, multiple access schemes play a critical role in providing the increasing demand in services for future terminals and applications. The current Long Term Evolution (LTE) [2] system is not able to efficiently support massive connectivity, especially in the uplink. Among the issues arise due to massive connectivity is the cost of signaling overhead and latency.

Sparse code multiple access (SCMA) is introduced in [3] as a new multiple access scheme. SCMA is a codebook-based non-orthogonal access mode allowing overloading of the system with a large number of SCMA layers to enable massive connectivity. Sparsity of SCMA codewords makes the near-optimal detection of over-laid SCMA layers practically feasible. In this paper, we employ SCMA with a contention based mechanism to address the problems described in terms of massive connectivity. Section II overviews the problems in current LTE systems to support uplink massive connectivity and describes the benefits of applying SCMA as a solution to existing issues. The system model for contention-based transmission is provided in Section III. Uplink contention based SCMA PHY and MAC layer designs are discussed in Section IV. Numerical results and analysis for a small packet transmission scenario are presented in Section V. The conclusion is given in Section VI.

## II. MOTIVATION AND OVERVIEW OF UL SCMA CONTENTION ACCESS

### A. Limitations of Existing LTE Technology

In LTE, uplink transmission is scheduled by a serving basestation with a request-grant procedure. A UE sends a scheduling request (SR) to the network in UL dedicated resources that occur periodically. The periodicity of such opportunity is typically every 5 or 10 ms. In the best case scenario, if data arrive at the UE right before the SR opportunity, it needs to wait for 7 ms between the request and uplink data transmission [2]. The delay is even larger if data arrive in between SR opportunities or the periodicity of SR is configured to be longer. This is disadvantageous for delay sensitive traffic of future applications.

The basestation, upon receiving an SR, performs scheduling for uplink data transmission. An uplink grant is then sent in the downlink control channel (PDCCH) [4]. The uplink grants, together with downlink resource assignments, occupy the downlink control region. The cost of such dynamic signalling for uplink is higher for small packets since the ratio of signalling overhead to useful payload is high. As an example, if the average number of control channel elements (CCEs) used for an uplink grant is 2, this is equivalent to 72 resource elements (REs) [5]. To transmit a small amount of data (e.g. 20 bytes of data using QPSK ½ occupying 160 REs), the UL grant signalling overhead (i.e. 72 REs) to the total resources occupied by the UL grant and UL payload (i.e. 72

REs + 160 REs) can be around 30%. Semi-persistent scheduling can be an option in order to reduce the dynamic signalling overhead [2]. However, such mechanism is more suitable for traffic arrival that exhibits some form of periodicity such as VoIP. Hence, it cannot efficiently support bursty traffic.

Conventional contention-based transmission mechanism, such as the schemes used in IEEE 802.11 standard [6], allows terminals to attempt to transmit data immediately after packet arrival without the need to wait for a transmission grant. However, it can only support one user data transmission in the wireless medium at any given time in the non-MIMO scenario. The adoption of multi-user MIMO schemes in 802.11 increases the simultaneous data connections in the network. However, more work is needed to support massive connectivity of 5G wireless networks.

To combat the problems described above, a new multiple-access technology is desired with the following features:

- Removal or reduction of the resource allocation and control signalling overhead,
- reduction of transmission latency, and
- support of large amount of users to enable massive connectivity.

*B. UL Contention Based Data Transmission Overview*

Physical random access channel (PRACH) [4] of LTE is used to setup a connection with the serving basestation and prepare for data transmission. Once the connection is established, the data transmission is carried out through a request and grant procedure described previously.

To reduce the latency and signaling overhead in UL, contention based data transmission can be used. UEs transmit data in such a way that they can contend for one or more resources without the need to go through the request and grant procedure. Such a scheme is a solution to some future applications with very stringent latency requirements.

Orthogonal frequency division multiple access (OFDMA) or single-carrier frequency division multiple access (SC-FDMA) can be used for either grant or grant-free UL data transmission. In the case of grant transmission mode, OFDMA/SC-FDMA resources are orthogonally split among the scheduled users. Users are informed about the scheduling decision through the control signaling at the cost of the extra overhead. In the case of the grant-less transmission mode, one or more predefined regions within the time-frequency plane of OFDMA or SC-FDMA is assigned to contention based transmission. The assigned time-frequency resources are shared among a pool of users which are allowed to use a specific region for contention transmission. Upon the arrival of a packet, a user attempts to transmit the data through a portion of the resources exist within a contention region. As multiple users share the same time-frequency resources, it might lead to non-orthogonal collision of the users' arriving signals at the receiver. As a result, the contention based OFDMA/SC-FDMA data transmissions incur significant performance degradations once the load of the system increases with a large number of active users.

The above problem is addressed by the proposed contention based SCMA which takes the overloading and non-orthogonality into account. The properties of an uplink contention based SCMA system are listed as follows:

- Same as OFDMA/SC-FDMA, one or a few predefined time-frequency regions are defined for the SCMA contention based transmission mode.
- Multiple-access is achievable through multiplexing of SCMA layers within a contention region. A layer is spread across the entire resources of a contention region.
- Every layer represents a user. A layer or user is characterized by a specific SCMA codebook. Furthermore, a user is represented with its corresponding contention region, SCMA codebook, and pilot sequence.
- SCMA codewords are sparse (in the sense of number of non-zero elements) such that a moderate-complexity multi-user detection based on the idea of message passing algorithm (MPA) is made possible [7][8]. MPA provides a near-optimal joint detection and cross-layer interference cancellation for SCMA.
- The system can be overloaded where the number of multiplexed layers is more than the length of codewords. The overloading feature of SCMA can provide massive connectivity with a limited complexity of detection thanks to the sparsity of SCMA codewords.

III. SYSTEM MODEL

*A. SCMA Multiple Access Transmission and Detection*

An SCMA encoder is defined as a map from $\log_2(M)$ bits to a $K$-dimensional complex codebook of size $M$ [3]. The $K$-dimensional complex codewords of a codebook are sparse vectors with $N < K$ non-zero entries. In the uplink contention based multiple access, a user is configured with a codebook. The user's data bits are mapped to a $K$-dimensional codeword selected from the codebook and transmitted on $K$ radio resources (e.g. OFDMA subcarriers). $K$ is the length of an SCMA codeword, or equivalently it represents the spreading factor of the system. Every SCMA block is carried over $K$ OFDMA tones. Depending on the size of a contention region, multiple non-overlapped SCMA blocks may fit within the assigned time-frequency resources.

Multiple access is achieved through the sharing of the same time-frequency resources among SCMA layers of multiple active users. Upon the reception of signals, multiple SCMA layers may collide with the same SCMA block. The number of simultaneous layers/users within a time slot varies depending on the traffic loading and possible packet retransmissions. Assume a time slot in which the signals of $U$ users arrive through different channels. Within an SCMA block, user $u$ ($u = 1,2,\ldots,U$) transmits codeword $\mathbf{x}_u$ selected from a codebook. The received signal after the synchronous layer multiplexing can be expressed as [3]

$$\mathbf{y} = \sum_{u=1}^{U} \sqrt{P_u}\,\text{diag}(\mathbf{h}_u)\mathbf{x}_u + \mathbf{i} \quad (1)$$

where **y** is the $K \times 1$ received vector over an SCMA block, $\mathbf{x}_u = [x_{1u}, x_{2u}, \ldots, x_{Ku}]^T$ is the vector of SCMA codeword of user $u$, $P_u$ is the received signal power of user $u$, $\mathbf{h}_u = [h_{1u}, h_{2u}, \ldots, h_{Ku}]^T$ is the channel vector of user $u$ over $K$ OFDMA tones of an SCMA block, $\text{diag}(\mathbf{h}_u)$ is a diagonal matrix where its $n$-th diagonal element is $h_{nu}$, and **i** is the ambient AWGN noise plus out-of-cell interference.

Due to the sparsity feature of SCMA codewords, the number of non-zero elements colliding at a give tone of **y** is effectively less than the total number of active layers ($U$). To take advantage of the sparsity and limit the complexity of detection, a MAP multi-user detection scheme is employed. In a typical scenario of UL transmission with 2 receive antennas, the complexity of MPA detection of 6 SCMA layers with 150% overloading is less than 4 times of a linear MMSE receiver. The details of the MPA detection can be found in [3] and the references within.

### B. SCMA Scalability for Massive Connectivity

The maximum number of codebooks, $J$, that can be generated is a function of $N$ (non-zero entries in a codeword) and $K$ (codeword length). Selection of $N$ non-zero positions within $K$ elements is simply a combination problem. The maximum number of such combinations is given by the binomial coefficient as follows

$$J = \binom{K}{N}. \qquad (2)$$

By multiplexing $J$ codewords from the codebooks over $K$ resources, the overloading factor (OF) is defined as:

$$\text{OF} = \frac{J}{K}. \qquad (3)$$

By adjusting the spreading factor, $K$, and the number of non-zero entries, $N$, different levels of overloading can be achieved with different number of codebooks. In order to support massive connectivity, it is desired to have an overloading factor much greater than 1.

As an example, for a 4-dimensional complex codebook ($K = 4$) with 2 non-zero entries ($N = 2$), a set of $J = 6$ codebooks can be generated from (2). An illustration is provided in Fig. 1. Two bits ($b_1, b_2$) from a data stream are mapped to a codeword. The data are then spread over 4 subcarriers. As expressed in (1), data streams of multiple users are overlaid with codewords from different codebooks.

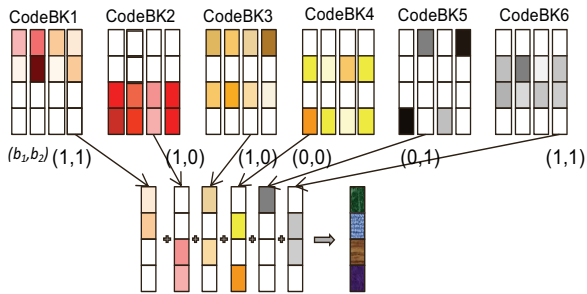

Fig. 1. SCMA codebooks, encoding, and multiplexing.

## IV. PHY AND MAC LAYER DESIGN

### A. UL SCMA Resource Definition

In order to support UL SCMA with contention based access, radio resources are defined for the proposed multiple access scheme. The basic resource for contention transmission is called a contention transmission unit (CTU). A combination of time, frequency, SCMA codebook, and pilot sequence defines a CTU as shown in Fig. 2. There are $J$ unique codebooks defined over a time-frequency resource. For each codebook, $L$ pilot sequences are associated with it. A total of $L \times J$ unique pilot sequences are defined. There are a total of $L \times J$ CTUs in the given time-frequency region.

Several terminals may share the same codebook, and due to the random access nature, those terminals may actively transmit at the same time. As codebooks go through different wireless channels, the MPA receiver is still able to detect the data streams even if they are carried over identical codebook as long as different pilot sequences are used. The receiver can estimate the channels of different terminals with the different pilots. Therefore, in a practical scenario, the number of active users ($U$) at a specific time slot can be potentially more than $J$. Codebook reuse can help to increase the effective overloading factor and the number of connections to realize the massive connectivity.

Consequently, user collision happens only if two or more users pick the same pilot sequence within a contention region. Pilot collisions need to be resolved via the random back-off mechanism as described in the next sub-section.

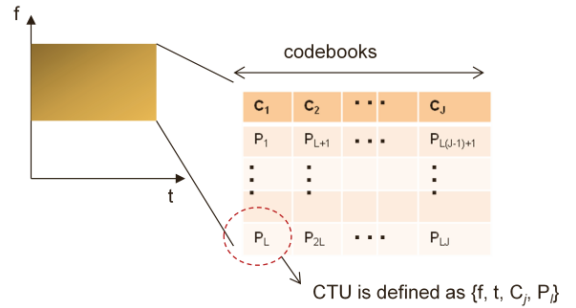

Fig. 2. Definition of a contention transmission unit (CTU).

The time-frequency resources on which the codebooks are overlaid form a contention region. This is shown in the designated time-frequency resources in Fig. 3. The size and the number of access regions depend on many factors such as the expected number of terminals and/or applications that are suitable for UL SCMA.

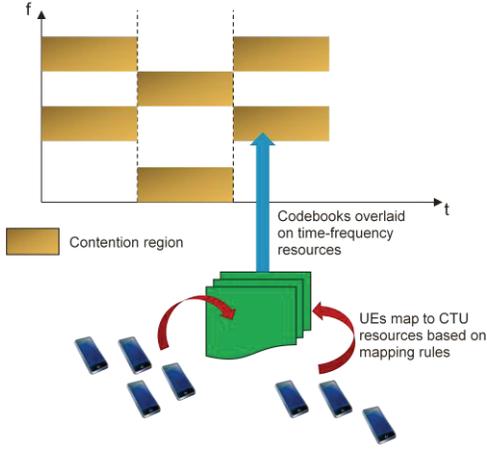

Fig. 3. Uplink contention-based regions within an OFDMA time-frequency plane.

### B. Transmission Mechanism

The proposed UL contention-based SCMA solution consists of the following two major components: i) a contention based multiple access mechanism with the possibility of multiple UEs contending for the same resource, and ii) non-adaptive transmission with predefined set(s) of coding and modulation levels.

Each UE transmits data in the predefined contention region with a CTU. The determination of the access region and CTU can be assigned explicitly by the network via semi-static signaling or implicitly derived from the UE ID. An example of a simple UE-to-CTU mapping rule can be CTU_index = UE_ID mod $N_{CTU}$, where the CTU index that a UE transmits its data over is a function of the UE ID and the total number of resources ($N_{CTU}$). This mapping rule can be changed in a predefined manner from time to time in order to provide collision diversity.

The multiple access mechanism allows contention to occur as multiple UEs may be assigned the same resource (CTU) as shown in Fig. 3. The network detects the uplink packets by attempting reception using all possible access codes assigned to the predefined contention region. Some blind detection and compressive sensing techniques can be used to perform joint data and activity detections [9]. In addition, cyclic redundancy check masked with UE ID or header identification may be used as detection criteria and enablers for HARQ retransmissions.

Uplink synchronization can be maintained with a timing advance procedure similar to existing LTE systems [2]. In the case where uplink timing alignment is lost and only traffic carried over contention-based SCMA is available, the timing advance procedure precedes actual data transmission. In a mixed traffic scenario, a UE can maintain uplink timing alignment by relying on traffic exchange in regular scheduled transmissions.

Since multiple UEs can be mapped to the same CTU for data transmission, collision can occur when more than one UE has data to transmit at the same time. Collision resolution can employ the random back-off procedure. Each UE picks a random back-off time (in units of transmission time interval (TTI)) from a back-off window and retransmit on the predefined CTU as the original transmission.

With the pre-defined contention regions and CTU assignments to UEs, and the blind detection techniques the dynamic UL grant is no longer necessary for contention based SCMA. Since UL grants occupy DL control channel resources, elimination of grant signaling for traffic with UL contention based SCMA means that DL signaling overhead can be reduced.

### V. NUMERICAL RESULTS AND ANALYSIS

In this section, we present some analysis on the scalability of SCMA and system-level simulation of the potential gain of SCMA over OFDMA contention-based access on the uplink.

### A. Scalability Analysis

A plot of the maximum number of codebooks $J$ that can be generated as a function of the spreading factor $K$ and the number of non-zero entries $N$ is shown in Fig. 4.

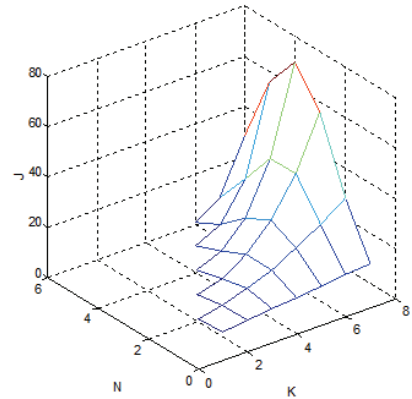

Fig. 4. Number of codebooks, $J$, as a function of $N$ and $K$.

It can be seen that by varying the spreading factor and the number of non-zero entries, the number of codebooks can be increased dramatically. As an example, when $K = 8, N = 4$, 70 codebooks can be generated. When more codebooks are generated, the number of CTU's in a contention region increases.

### B. Simulation Setup and Assumptions

System-level simulation for a small packet application scenario is performed on a 19-cell 3-sector network. Evaluation is done on the 2GHz carrier frequency. It should be noted that existing cellular bands will be part of the 5G system that includes licensed, unlicensed and high frequency spectrum. Different numbers of users are dropped randomly in the network and statistics are collected for a sector in the centre cell. Uplink traffic for each user follows a Poisson distribution with a mean packet inter-arrival time of 160 ms per user. A specific traffic load in each sector is obtained by configuring a different number of active users in the sector. A contention region with a size of 4 LTE resource block (RB) pairs are simulated and flat Rayleigh fading channels are modeled. An RB pair is equivalent to 12 subcarriers in frequency and 14 OFDM symbols in time [2]. For fair comparison of the SCMA and OFDMA schemes, same data size that fits into the equivalent of one RB pair OFDMA resource is considered for each transmission with a fixed spectral efficiency of 1 bit/s/Hz. The data transmitted in SCMA is spread over the 4 RB pairs. For OFDMA, it is transmitted in one of the 4 RB pairs. Open-loop power control for LTE [4] is applied to determine the transmission power at the user terminal with fixed MCS. As a result, the interference from the neighboring sectors to the center sector can be captured, which varies in time. The MPA receiver is employed for SCMA [3] to handle non-orthogonal signal processing. A linear MMSE receiver is employed for OFDMA as a baseline LTE scenario.

We consider an application scenario for small packet transmission with low latency. Given a latency constraint, a packet is dropped when the waiting and transmission time is beyond the latency requirement. Perfect channel estimation is assumed in the simulation. The detailed simulation assumptions are described in Table I.

TABLE I. SIMULATION ASSUMPTIONS

| Parameters | Settings |
| --- | --- |
| Network layout | 19 cells with 3 sectors per cell |
| Inter-site distance | 500 m |
| Channel | Flat Rayleigh fading |
| Antenna configuration | SIMO 1x2, uncorrelated antennas |
| Modulation and coding | Fixed. QPSK ½ for OFDMA; Codebook with $K = 4$, $N = 2$, $J = 6$, code rate ½ for SCMA |
| Channel estimation | Perfect |
| Traffic pattern | Poisson packet arrival with a mean of 160 ms configurable number of active users in each sector |
| UE max Tx power | 23 dBm |
| Power control [4] | $\alpha = 1.0$ (full pathloss compensation), $P_0 = -95$ dBm |
| Receiver model | MPA receiver for SCMA; linear MMSE receiver for OFDMA |

## C. Performance and analysis

The UL contention based OFDMA and SCMA schemes are evaluated for transmission of small packets with low latency requirements such that the delay is within 5 ms interval. Due to the tight latency requirement, no retransmission opportunity is allowed.

Fig. 5 demonstrates the distribution of user packet drop rate (or loss) as a function of traffic loading for contention based SCMA and contention based OFDMA. By definition, a packet is lost if its transmission fails at the first attempt. It is shown clearly that the OFDMA packet drop rate performance degrades much faster with load increasing than SCMA. SCMA has a lower user packet drop rate distributions than OFDMA for various traffic loadings.

In the following, the system performance is evaluated based on outage criteria. The system operates in a traffic load to satisfy the QoS conditions defined by both UE outage and network outage. A UE is in outage if its packet drop rate is more than 2%, and the system outage is defined by the percentage of the users in outage.

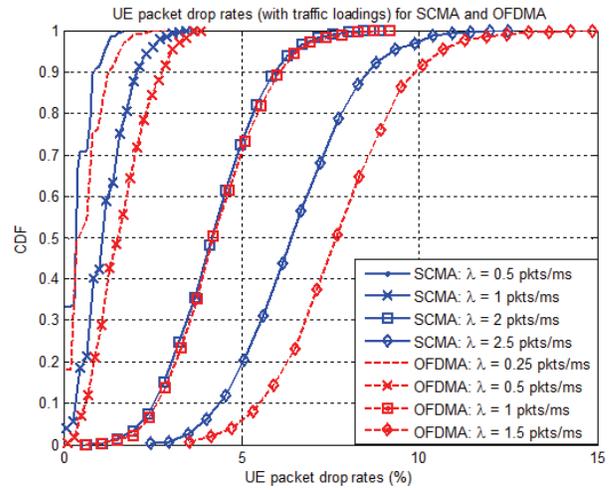

Fig. 5. CDF of user packet drop rate as a function of traffic loading for SCMA and OFDMA.

It is found that SCMA and OFDMA perform in different operating points in terms of traffic load to achieve the same QoS. Specifically, SCMA demonstrates a large advantage over OFDMA in terms of supported number of active users to achieve the same system outage. For example, to achieve a system outage of 2%, SCMA can support 119 users while OFDM support 42 users. Fig. 6 shows the system capacity per sector for SCMA and OFDMA when they achieve the system outage of 2% and 5%, respectively. As shown in Fig. 7, the simulated SCMA gains over OFDMA are 2.83 and 2.74 in terms of number of supported users for given system outages of 2% and 5%, respectively.

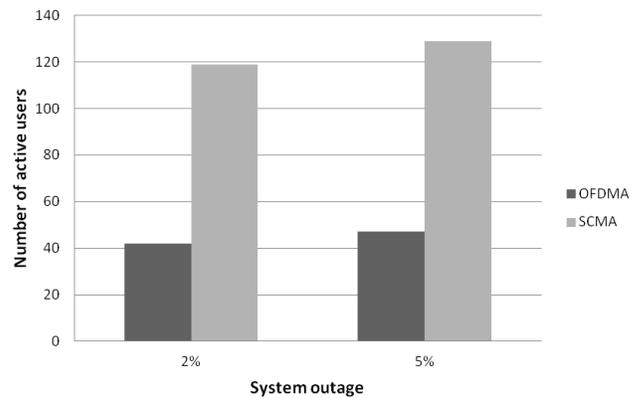

Fig. 6. Capacity supported based on different system outages.

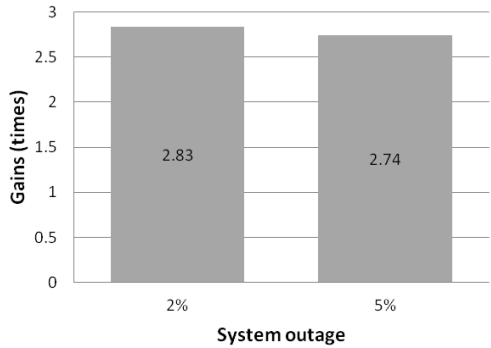

Fig. 7. Outage capacity gain of SCMA over OFDMA.

*D. PAPR Reduction for SCMA*

One issue related to OFDMA signal is its high range of peak-to-average power ratio (PAPR). SC-FDMA is an alternative solution adopted by UL LTE to reduce PAPR which is more suitable for low cost terminals. As mentioned before, SCMA codewords are carried over OFDMA tones. Hence, the problem of high PAPR can inherently pass to SCMA-OFDM transmission mode as well. As opposed to original OFDM, in SCMA-OFDM there is an extra degree of freedom in which the issue of PAPR can be addressed during the process of the SCMA codebook design. The general method for SCMA codebook design is introduced in [10]. Following this method, low projection codebooks are designed such that the outcome guarantees low PAPR (<5 dB) for SCMA-OFDM transmission in certain narrow band scenarios. This is comparable with current low PAPR SC-FDMA waveform.

## VI. CONCLUSION

In this paper, an uplink contention-based SCMA scheme is proposed. The SCMA parameters can be adjusted to provide different levels of overloading, thus suitable to meet a diverse traffic connectivity requirement of 5G wireless networks. We describe the system design aspects of the new multiple access scheme. System-level evaluations of a small packet application scenario are provided for contention-based UL SCMA. SCMA is compared to OFDMA in terms of connectivity and drop rate under a tight latency requirement. The simulation results demonstrate the potential gain of contention-based SCMA over contention-based OFDMA scheme for small packet transmission. SCMA can provide around 2.8 times gain over OFDMA in terms of supported active users in a contention based system with low latency traffic.

Therefore, the uplink contention-based SCMA scheme can be a promising technology for 5G wireless networks for data transmission with low signaling overhead, low delay, and support of massive connectivity.